\documentclass[12pt,draftclsnofoot,onecolumn]{IEEEtran}

\usepackage{cite}      

\usepackage{graphicx}  

\usepackage{psfrag}    
\usepackage{amssymb}
\newcommand{\real}{{\mathbb{R}}}

\def\ns{\hspace{-1mm}}

\def\gB{{\cal B}}

\def\gJ{{\cal J}}

\def\gR{{\cal R}}
\def\gS{{\cal S}}

\def\gV{{\cal V}}

\def\gX{{\cal X}}

\def\gZ{{\cal Z}}
\begin{document}
%
\title{\Large 
Geometric Methods for Invariant-Zero Cancellation in Linear Multivariable Systems: 
Illustrative Examples}
%
%
%
\author{Elena~Zattoni 
\thanks{The author is with the Department of Electrical, Electronic and Information
Engineering ``Guglielmo Marconi'', Alma Mater Studiorum~$\cdot$~University of Bologna, 
Viale Risorgimento~2, 40136 Bologna, Italy. E-mail: elena.zattoni@unibo.it}
}

\maketitle
\begin{abstract}
\noindent
This note presents some numerical examples worked out in order to show the reader
how to implement, within a widely accessible computational setting, the methodology 
for achieving zero cancellation in linear multivariable systems discussed in 
\cite{Zattoni-2013}. The results are evaluated in light of applicability and 
performance of different methods available in the literature.
\end{abstract}


%
\IEEEpeerreviewmaketitle

\thispagestyle{empty}
%
%
\section{Introduction}
The methodology for achieving zero cancellation in linear multivariable systems developed 
in \cite{Zattoni-2013} is based on the geometric characterization of the invariant zeros
of a linear time-invariant multivariable system as the internal unassignable eigenvalues 
of the maximal output nulling controlled invariant subspace \cite{Basile-M-1992}. In 
particular, in \cite{Zattoni-2013} it is shown that a series of three state-space basis 
transformations, determined in connection with the maximal output nulling controlled 
invariant subspace and a friend linear map, results in a representation of the system 
where the structure of the minimum-phase invariant zeros is caught by a certain pair of 
matrices. Hence, the linear maps associated with those matrices are used to define the 
feedforward compensator achieving zero cancellation while retaining some special properties 
of the original system, such as stabilizability and right-invertibility. 
\par
In this note, three numerical examples are presented. The first example is worked out in 
detail, with the aim of illustrating every single step of the application of the procedure 
proposed in \cite{Zattoni-2013}. The second example is borrowed from the literature, with 
the purpose of comparing the results obtained with the techniques described in 
\cite{Zattoni-2013} to those provided by the approach introduced in \cite{Wan-RS-2009}. As
to the third example, it has been known for a long time that zero assignment can effectively 
be exploited to improve the transient response of a control system \cite{EmamiNaeini-F-1982}. 
Thus, the third example is devised in order to show how a slight modification of the method 
discussed in \cite{Zattoni-2013} leads to a feedforward compensator that not only attains 
zero cancellation with the additional constraints of retaining reachability and 
right-invertibility, but also ameliorates the system step response by eliminating the 
overshoot.
\par
The computational framework consists of the Matlab files implementing the geometric approach 
algorithms, first appeared with \cite{Basile-M-1992} and now available online in an upgraded 
version. The variables are displayed in scaled fixed point format with five digits, although 
computations are made in floating point precision.
\par
{\em Notation:}
$\real$ stands for the set of real numbers. Matrices and linear maps are denoted by 
upper-case letters, like $A$. The spectrum, the image, and the kernel of~$A$ are 
denoted by $\sigma(A)$, ${\rm im}\,A$, and ${\rm ker}\,A$, respectively. Vector 
spaces and subspaces are denoted by calligraphic letters, like $\gV$. Vector spaces
and subspaces are further characterized by subscripts when the same subscripts are
used to denote the matrices of the systems they refer to. The symbols $I_n$ and 
$O_{m\times n}$ are respectively used for the identity matrix of dimension $n$ 
and the $m\,{\times}\,n$ zero matrix (subscripts are omitted when the dimensions 
can be inferred from the context). 
%
%
\section{Example~1}
\label{sect_example1}
Consider the continuous-time linear time-invariant system 
\begin{eqnarray}
\dot x(t) \ns&\ns = \ns&\ns A\,x(t)+B\,u(t),\label{eq_sys1}\\
y(t)      \ns&\ns = \ns&\ns C\,x(t)+D\,u(t),\label{eq_sys2}
\end{eqnarray}
and assume
\begin{eqnarray*}
A\ns&\ns=\ns&\ns\left[\begin{array}{rrrrr}
   -0.79	&   -1.89	&   -1	&   -1.01	&   -0.2\\
    0.89	&   -4.3	&   -0.76	&   -0.48	&   -0.12\\
    0.8	&   -5.57	&   -3.25	&   -3.01	&   -1.52\\
   -1.18	&    3.41	&    0.26	&   -1.03	&    0.92\\
    1.62	&   -6.41	&    0.55	&   -4.15	&   -5.63
\end{array}\right],\quad
B = \left[\begin{array}{rrrr}
     1	&     0	&     0	&     0\\
     0	&     1     &	0     &	1\\
     2	&     1     &	0     &	0\\
     1    	&	-1    &	0     &	0\\
    -1     	&	0     &	2     &	0
\end{array}\right],\\
C\ns&\ns=\ns&\ns\left[\begin{array}{rrrrr}
    1		&        0	&         0	&         0	&	0\\	
    0.89	&   	-2.8	&   	-0.76	&   	-0.48	&	-1.12\\
   -0.29	&   	-0.89	&   	-0.25	&   	-1.51 &  	-0.2
\end{array}\right],\quad
D = \left[\begin{array}{rrrr}
     0	&     0	&     0	&     0\\
     0     	&	1     &	0     &	0\\
     1     	&	0     &	0     &	0
\end{array}\right].
\end{eqnarray*}
The matrices of the given system satisfy the rank conditions stated in 
\cite[Section~II]{Zattoni-2013}. Note that the given system is reachable, 
since the minimal $A$-invariant subspace containing $\gB\,{=}\,{\rm im}\,B$
or, equivalently, the reachable subspace of the pair $(A,B)$ is given by
$\gR\,{=}\,{\rm min\,}\gJ(A,\gB)\,{=}\,\gX\,{=}\,\real^5$. Moreover, the 
given system is right-invertible, since the maximal output-nulling controlled
invariant subspace $\gV^\ast$ and the minimal input-containing conditioned 
invariant subspace $\gS^\ast$ are respectively given by
\[
\gV^\ast={\rm im}\,
\left[\begin{array}{rrr}
	 0	&   	 0	&   	0\\
    	-0.6695	&    	 0	&    	0\\
   	 0.6180	&    	-0.5547	&   	0\\
    	-0.4120	&    	-0.8321	&    	0\\
    	 0	&    	 0	&   	-1
\end{array}\right],\quad
\gS^\ast={\rm im}\,
\left[\begin{array}{rrr}
    0.4862	&   0		&	0\\
   	0	&   -1	&	0\\
    0.4813	&   0		&	0\\
   -0.7293	&   0		&	0\\
    	0	&   0		&	1
\end{array}\right],
\]
and, therefore, $\gV^\ast\,{+}\,\gS^\ast\,{=}\,\gX$. Furthermore, the set of the invariant 
zeros $\gZ(A,B,C,D)$, which is the union of the sets of the minimum-phase invariant zeros 
$\gZ_{MP}(A,B,C,D)$ and of the nonminimum-phase invariant zeros $\gZ_{NMP}(A,B,C,D)$, is 
given by
\[
\gZ(A,B,C,D)\,{=}\,\gZ_{MP}(A,B,C,D)\,{\cup}\,\gZ_{NMP}(A,B,C,D)
\,{=}\,\{-1.2509\}\,{\cup}\,\{0.7534\}.
\]
On these conditions, the geometric methodology presented in \cite{Zattoni-2013} allows us 
to design a feedforward compensator that cancels the minimum-phase zero $z_{MP}\,{=}\,{-}
\,1.2509$, while maintaining reachability and right-invertibility in the resulting cascade 
system.
\par
The first step of the geometric method requires us to pick a linear map $F$ such that 
$(A\,{+}\,BF)\,\gV^\ast\,{\subseteq}\,\gV^\ast$ and $\gV^\ast\,{\subseteq}\,{\rm ker}
\,(C\,{+}\,DF)$. A linear map $F$ satisfying these conditions is represented by the 
matrix
\[
F=\left[\begin{array}{rrrrr}
	0	&    0.7121	&	0.1166	&	1.5990	&    0.2000\\
	0	&    1.0731	&	-0.5352	&	1.3434	&    1.1200\\
	0	&    1.9832	&	-0.8226	&	2.7324	&    2.3320\\
	0	&    0.6742	&	-0.3712	&	0.7916	&   -1.0000
\end{array}\right].
\]
In order to determine the first similarity transformation $T$, defined according to
\cite[Lemma~1]{Zattoni-2013}, it is worth noting that 
\[
\gR_{\gV^\ast}\,{=}\,\gV^\ast\,{\cap}\,\gS^\ast\,{=}\,
{\rm im}\,\left[\begin{array}{c}0 \\ 0 \\ 0 \\ 0 \\ 1
\end{array}\right].
\] 
Then,
\[
T=\left[\begin{array}{r|rr|rr}
	0	&	0		&	0	&   -0.4862	&	0\\
	0	&	0.6695	&	0	&    0	&	-1\\
	0	&	-0.6180	&   -0.5547	&   -0.4813	&	0\\
	0	&	0.4120	&   -0.8321	&    0.7293	&	0\\
	1	&	0		&	0	&    0    	&	0
\end{array}\right].
\]
Consequently, the matrices $A_F'$ and $C_F'$, defined according to 
\cite[eqs.~(7),\,(8)]{Zattoni-2013}, are
\begin{eqnarray*}
A_F'\ns&\ns=\ns&\ns\left[\begin{array}{r|rr|rr}
   -1.1660	&   -1.4810	&    0.9086	&   -0.4116	&    3.1557\\ \hline
	0	&    0.0040	&   -0.6763	&   -0.2900	&   -3.0932\\
	0	&   -1.3907	&   -0.5015	&   -1.8603	&    0.0092\\ \hline
	0	&    0	&    0	&   -2.5481	&   -2.4227\\
	0	&    0	&    0	&   -1.7707 & -4.6237
\end{array}\right],\\
C_F'\ns&\ns=\ns&\ns\left[\begin{array}{r|rr|rr}
	0	&	0	&	0	&   -0.4862	&	0\\
	0	&	0	&	0	&    0.8204	&    1.7269\\
	0	&	0	&	0	&    0.2701	&    0.1779
\end{array}\right].\\
\end{eqnarray*}
According to \cite[Lemma~2]{Zattoni-2013}, the similarity transformation $T'$ is 
determined by solving the Sylvester equation. The relation
\begin{eqnarray*}
\sigma(A_{11}')\cap\sigma(A_{22}')=
\{-1.1660\}\cap\{0.7534,\,-1.2509\}=\emptyset
\end{eqnarray*}
ensures the existence and uniqueness of the solution of the Sylvester equation: i.e.,
\[
X=\left[\begin{array}{rr}
-1.7146 & -0.3776
\end{array}\right].
\]
Hence, 
\[
T'=\left[\begin{array}{r|rr|rr}
1	&	-1.7146		& -0.3776	& 0	&	 0\\ \hline
0	&	 1		&	0	& 0	&	 0\\
0	&	 0		&	1	& 0	&	 0\\ \hline
0	&	 0		&	0	& 1	&	 0\\
0	&	 0		&	0	& 0	&	 1
\end{array}\right].
\]
Consequently, $A_F''$ and $C_F''$, defined according to \cite[eqs.~(13),\,(14)]{Zattoni-2013},
are
\[
A_F''=\left[\begin{array}{r|rr|rr}
   -1.1660	&    0	&    0	&   -1.6113	&   -2.1444\\ \hline
	0	&    0.0040	&   -0.6763	&   -0.2900	&   -3.0932\\
	0	&   -1.3907	&   -0.5015	&   -1.8603	&    0.0092\\ \hline
	0	&    0	&    0	&   -2.5481	&   -2.4227\\
	0	&    0	&    0	&   -1.7707 & -4.6237
\end{array}\right]
\]
and $C_F''\,{=}\,C_F'$. In order to determine the similarity transformation $T''$, considered
in \cite[Lemma~3]{Zattoni-2013}, the matrix 
\[
J'=\left[\begin{array}{rr}
J_S & J_U\end{array}\right]=
\left[\begin{array}{r|r}
0.4744 & 0.6699\\
0.8803 & -0.7424
\end{array}\right],
\]
is computed. Thus,
\[
T''=\left[\begin{array}{r|rr|rr}
1	&	 0		& 	0	& 0	&	 0\\ \hline
0	&	 0.4744		&	0.6699	& 0	&	 0\\
0	&	 0.8803 	& 	-0.7424	& 0	&	 0\\ \hline
0	&	 0		&	0	& 1	&	 0\\
0	&	 0		&	0	& 0	&	 1
\end{array}\right].
\]
As a consequence, $A_F'''$ and $C_F'''$, defined according to \cite[eqs.~(15),\,(16)]{Zattoni-2013}, 
are
\[
A_F'''=\left[\begin{array}{r|rr|rr}
   -1.1660	&    0	&    0	&   -1.6113	&   -2.1444\\ \hline
	0	&   -1.2509	&    0	&   -1.5517	&   -2.4314\\
	0	&    0	&    0.7534	&    0.6658	&   -2.8954\\ \hline
	0	&    0	&    0	&   -2.5481	&   -2.4227\\
	0	&    0	&    0	&   -1.7707 &   -4.6237
\end{array}\right]
\]
and $C_F'''\,{=}\,C_F''$. Then, one gets the resolving subspace
\[
\gV_S^\ast={\rm im}\,{V_S^\ast}'''
={\rm im}\,
\left[\begin{array}{r}
	0\\ \hline
   	1\\
    	0\\ \hline
   	0\\
    	0
\end{array}\right],
\]
according to \cite[Theorem~1]{Zattoni-2013}. With respect to the original coordinates, 
$\gV_S^\ast$ is given by
\[
\gV_S^\ast={\rm im}\,\left(\bar T\,{V_S^\ast}'''\right)={\rm im}\,
\left[\begin{array}{r}
	0\\
   	0.3176\\
     -0.7815\\
     -0.5370\\
     -1.1458
\end{array}\right],
\]
where $\bar T=T\,T'\,T''$. According to \cite[Corollary~1]{Zattoni-2013}, from $A_F'''$, $F$, 
and $V_S^\ast$ one also gets the matrices that point out the structure of the minimum-phase 
invariant zeros of the original system: i.e.,
\[
W=-1.2509,\quad
L=\left[\begin{array}{r}
-0.9528\\
-1.2457\\
-2.8665\\
1.2249
\end{array}\right].
\]
Therefore, the matrices of the feedforward compensator, defined according to 
\cite[eqs.~(24),\,(25)]{Zattoni-2013}, are
\begin{eqnarray*}
A_f\ns&\ns=\ns&\ns-1.2509,\hspace{15mm}
B_f = \left[\begin{array}{r|rrrr}
1 & 0 & 0 & 0 & 0
\end{array}\right],\\
C_f\ns&\ns=\ns&\ns\left[\begin{array}{r}
-0.9528\\
-1.2457\\
-2.8665\\
1.2249
\end{array}\right],\quad
D_f = \left[\begin{array}{r|rrrr}
0 & 1 & 0 & 0 & 0\\
0 & 0 & 1 & 0 & 0\\
0 & 0 & 0 & 1 & 0\\
0 & 0 & 0 & 0 & 1
\end{array}\right].
\end{eqnarray*}
The matrices of the cascade in \cite[eqs.~(5),\,(6)]{Zattoni-2013},
defined according to \cite[eqs.~(32),\,(33)]{Zattoni-2013}, are
\begin{eqnarray*}
A_e\ns&\ns=\ns&\ns\left[\begin{array}{rrrrr}
   -0.79	&   -1.89	&   -1		&   -1.01	&   -0.2\\
    0.89	&   -4.3	&   -0.76	&   -0.48	&   -0.12\\
    0.8		&   -5.57	&   -3.25	&   -3.01	&   -1.52\\
   -1.18	&    3.41	&    0.26	&   -1.03	&    0.92\\
    1.62	&   -6.41	&    0.55	&   -4.15	&   -5.63
\end{array}\right],\quad
B_e=\left[\begin{array}{r|rrrr}
0		&     	 1	&	0	&     	0	&     0\\
-0.3176     	&	 0	&	1	&	0     	&     1\\
0.7815		&	 2	&	1	&	0	&     0\\
0.5370		&	 1    	&	-1      &	0     	&     0\\
1.1458		&	-1     	&	0       &	2       &     0
\end{array}\right],\\
C_e\ns&\ns=\ns&\ns\left[\begin{array}{rrrrr}
    1		&        0	&         0	&         0	&	0\\	
    0.89	&   	-2.8	&   	-0.76	&   	-0.48	&	-1.12\\
   -0.29	&   	-0.89	&   	-0.25	&   	-1.51 &  	-0.2
\end{array}\right],\quad
D_e = \left[\begin{array}{r|rrrr}
0&     0	&     0	&     0	&     0\\
0&     0     	&	1     &	0     &	0\\
0&     1     	&	0     &	0     &	0
\end{array}\right].
\end{eqnarray*}
Note that the cascade is reachable, since $\gR_e\,{=}\,{\rm min\,}\gJ\,(A_e,\gB_e)\,{=}\,\gX$,
which means that the feedforward compensator has maintained reachability. The cascade is 
right-invertible, since 
\[
\gV_e^\ast={\rm im}\,
\left[\begin{array}{rrr}
	 0	&   	 0	&   	0\\
    	-0.6695	&    	 0	&    	0\\
   	 0.6180	&    	-0.5547	&   	0\\
    	-0.4120	&    	-0.8321	&    	0\\
    	 0	&    	 0	&   	-1
\end{array}\right],\quad
\gS_e^\ast={\rm im}\,
\left[\begin{array}{rrrr}
    0.4863	&   0		&	0	& 0\\
   	0	&   -1		&	0	& 0\\
    0.4949	&   0		&	0.8242  & 0\\
   -0.7202	&   0		&	0.5663	& 0\\
    	0	&   0		&	0	& 1
\end{array}\right],
\]
and, therefore, $\gV_e^\ast\,{+}\,\gS_e^\ast\,{=}\,\gX$. This shows that the precompensator has
kept right-invertibility. Furthermore, the set of the invariant zeros is 
$\gZ(A_e,B_e,C_e,D_e)\,{=}\,\{0.7534\}$,
which means that the minimum-phase invariant zero of the original system has been cancelled, while 
the nonminimum-phase invariant zero is still present.
%
%
\section{Example~2}
\label{sect_example2}
Consider the continuous-time linear time-invariant system (\ref{eq_sys1}),\,(\ref{eq_sys2}) 
and assume
\begin{eqnarray*}
A\ns&\ns=\ns&\ns\left[\begin{array}{rrrrrrr}
   -1	&    1	&    0	&    1	&    2	& 0	& 0\\
    0	&   -1	&    0	&    2	&    1	& 0	& 0\\
    0	&    0	&   -1	&    0	&    0	& 0	& 0\\
    0	&    0	&    0	&    1	&    0	& 0	& 0\\
    0	&    0	&    0	&    0	&    1	& 0	& 0\\
    0	&    1	&    0	&    1	&    0	& 1	& 1\\
    0	&    0	&    1	&    0	&    1	& 0	& 0
\end{array}\right],\quad
B = \left[\begin{array}{rr}
     0		&     0	\\
     0		&     0	\\
     0		&     0	\\
     0		&     0	\\
     0		&     0	\\
     1		&     0	\\
     0		&     1	
\end{array}\right],\\
C\ns&\ns=\ns&\ns\left[\begin{array}{rrrrrrr}
    0	&  0	&  0	&  1	& 0	& 0 & 0\\
    0	&  0	&  0	&  0	& 1	& 0 & 0
\end{array}\right],\hspace{14mm}
D = \left[\begin{array}{rr}
     0		&     0	\\
     0     	&     0 
\end{array}\right].
\end{eqnarray*}
The matrices of the system satisfy the rank conditions stated in \cite[Section~II]{Zattoni-2013}.
The system is not reachable and not right-invertible. The system has an invariant zero at $-1$
with multiplicity equal to $3$. By applying the standard procedure, one gets that the subspace 
$\gV_S^\ast$ is given by
\[
\gV_S^\ast={\rm im}\,V_S^\ast={\rm im}\,
\left[\begin{array}{rrr}
	-1	&	0	&	0\\
	 0	&	-1	&	0\\
	 0	&	0	&	1\\
	 0	&	0	&	0\\
	 0	&	0	&	0\\
	 0	&	0.3333	&	0.1667\\
	 0	&	0	&	-0.5\\
\end{array}\right],
\]
and the structure of the minimum-phase invariant zero is given by the matrices
\[
W=\left[\begin{array}{rrr}
	-1	&  1	& 0\\
	 0	& -1	& 0\\
	 0	&  0	& -1
\end{array}\right],\quad
L=\left[\begin{array}{rrr}
	0	& 0.3333 	& 0.1667\\
	0	& 0	 	& -0.5
\end{array}\right].
\]
In particular, the matrix $W$ shows that the zero dynamics have a defective eigenvalue. 
Nonetheless, by applying the procedure described in \cite{Zattoni-2013}, one gets the 
following matrices for the feedforward compensator
\begin{eqnarray*}
A_f\ns&\ns=\ns&\ns W,\quad
B_f = 	\left[\begin{array}{rr}
		I_3 & O_{3\times 2}
	\end{array}\right],\quad
C_f=	L,\quad
D_f = 	\left[\begin{array}{rr}
		O_{2\times 3} & I_2 
\end{array}\right].
\end{eqnarray*}
The matrices of the equivalent form of the cascade are
\begin{eqnarray*}
A_e\ns&\ns=\ns&\ns A,\quad
B_e=\left[\begin{array}{rr}
		-V_S^\ast & B
\end{array}\right],\quad
C_e= C,\quad
D_e = O_{2\times 5}.
\end{eqnarray*}
The system thus obtained is not reachable, is not right-invertible and does not have any invariant
zeros. It is worth noting that, in comparison with the solution proposed in \cite{Wan-RS-2009},
where the feedforward compensator has dynamic order equal to $6$, the solution proposed herein 
guarantees the feedforward compensator with the minimal dynamic order. In the specific case, the 
dynamic order of the feedforward compensator is $3$.
%
%
\section{Example~3}
\label{sect_example3}
\begin{figure}
\begin{center}
\includegraphics[width=1.1\columnwidth]{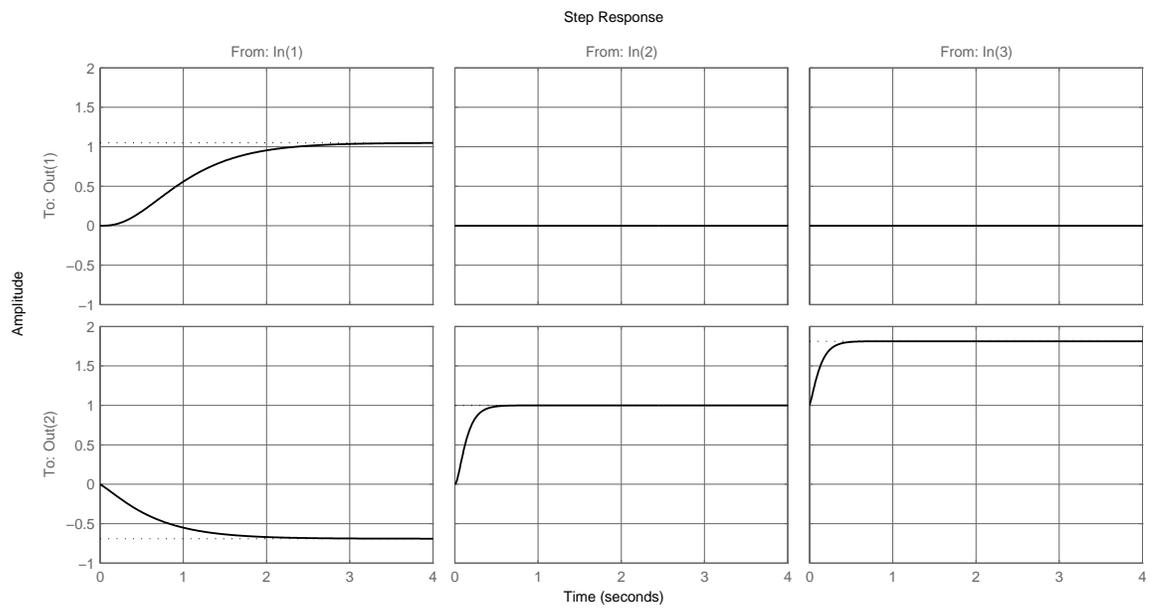}
\caption{Example~3: Step response of the equivalent cascade system}
\label{fig_overshoot2}
\end{center}
\vspace{-5mm}
\end{figure}

Consider the continuous-time linear time-invariant system (\ref{eq_sys1}),\,(\ref{eq_sys2})
and assume
\begin{eqnarray*}
A\ns&\ns=\ns&\ns\left[\begin{array}{rrrrr}
   -4	&    4	&    1	&    0	&    0\\
    0	&   -5	&   -3	&    0	&    0\\
    0	&    2	&    0	&    0	&    0\\
    0	&    0	&    0	&   -30	&   -12.5\\
    0	&    0	&    0	&    16	&    0
\end{array}\right],\quad
B = \left[\begin{array}{rrrr}
     0		&     0		&     	0\\
     3.4640	&     0     	&	0\\
     0		&     0     	&	0\\
     0    	&     3.5355    &	1\\
     0     	&     0     	&	1
\end{array}\right],\\
C\ns&\ns=\ns&\ns\left[\begin{array}{rrrrr}
    3.4640	&        0	&         0	&         0	&	0\\	
    0		&   	 0	&   	  1	&   	  0	&	3.5355
\end{array}\right],\hspace{10mm}
D = \left[\begin{array}{rrrr}
     0		&     0	&     0\\
    -0.5     	&     0 &     1
\end{array}\right].
\end{eqnarray*}
The matrices of the system satisfy the rank conditions stated in \cite[Section~II]{Zattoni-2013}.
The system is reachable and right-invertible. The system has only one zero, which is minimum-phase:
namely, $z_{MP}\,{=}\,-0.5$. 
By applying the standard procedure, one gets that the subspace $\gV_S^\ast$ is given by
\[
\gV_S^\ast={\rm im}\,
\left[\begin{array}{r}
	0\\
     -0.2425\\
      0.9701\\
      0.0281\\
     -0.0849
\end{array}\right],
\]
and the structure of the minimum-phase invariant zero is given by the matrices
\[
W=-0.5,\quad
L=\left[\begin{array}{r}
	0.5251\\
	0.0497\\
       -0.4074
\end{array}\right].
\]
A slight modification of the feedforward compensator design, carried out according to 
\cite[Remark~4]{Zattoni-2013}, yields the feedforward compensator matrices
\begin{eqnarray*}
A_f\ns&\ns=\ns&\ns-0.5,\quad
B_f = \left[\begin{array}{r|rr}
1 & 0 & 0 
\end{array}\right],\quad
C_f=\left[\begin{array}{r}
	0.5251\\
	0.0497\\
       -0.4074
\end{array}\right],\quad
D_f = \left[\begin{array}{r|rr}
0 &  0 & 0 \\
0 &  1 & 0 \\
0 &  0 & 1 
\end{array}\right].
\end{eqnarray*}
Consequently, the matrices of the equivalent form of the cascade system are
\begin{eqnarray*}
A_e\ns&\ns=\ns&\ns\left[\begin{array}{rrrrr}
   -4	&    4	&    1	&    0	&    0\\
    0	&   -5	&   -3	&    0	&    0\\
    0	&    2	&    0	&    0	&    0\\
    0	&    0	&    0	&   -30	&   -12.5\\
    0	&    0	&    0	&    16	&    0
\end{array}\right],\quad
B_e = \left[\begin{array}{r|rr}
	0	& 0	& 0\\
     	0.2425	& 0	& 0\\
      	-0.9701 & 0	& 0\\
      	-0.0281 & 3.5355& 1\\
     	0.0849	& 0	& 1
\end{array}\right],\\
C_e\ns&\ns=\ns&\ns\left[\begin{array}{rrrrr}
    3.4640	&        0	&         0	&         0	&	0\\	
    0		&   	 0	&   	  1	&   	  0	&	3.5355
\end{array}\right],\hspace{10mm}
D_e= \left[\begin{array}{r|rr}
     0		&     0	&     0\\
     0     	&     0 &     1
\end{array}\right].
\end{eqnarray*}
%
%
\begin{figure}
\begin{center}
\includegraphics[width=1.1\columnwidth]{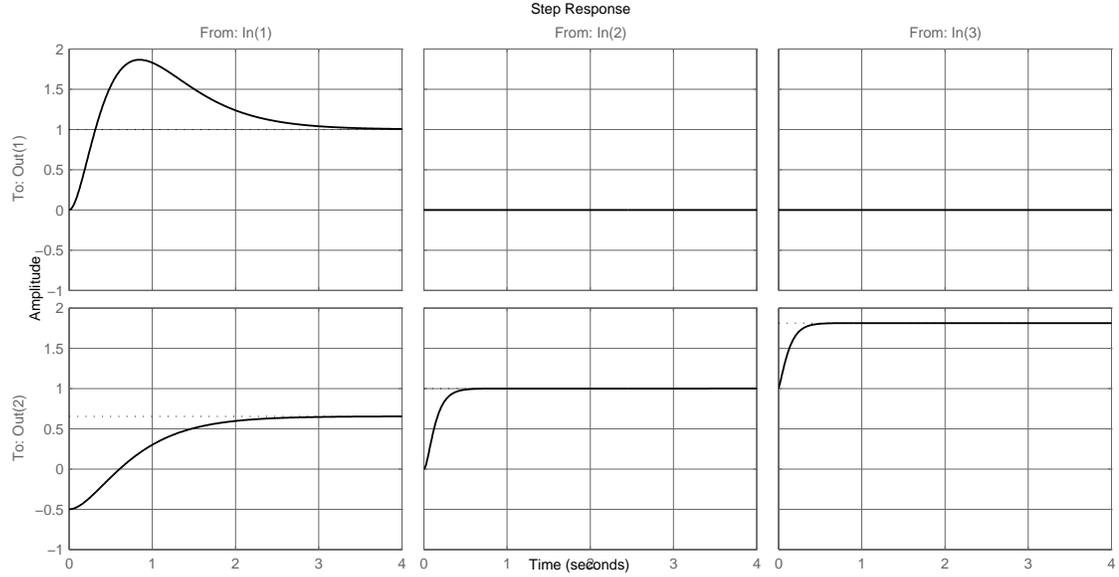}
\caption{Example~3: Step response of the original system}
\label{fig_overshoot1}
\end{center}
\vspace{-5mm}
\end{figure}
%
%
The cascade system thus obtained is reachable, right-invertible and without invariant zeros.
The comparison between the step response of the original system and that of the cascade shows
that zero cancellation has yielded the elimination of the overshoot in the new system 
(Figures~\ref{fig_overshoot1} and \ref{fig_overshoot2}). However, as mentioned in 
\cite[Remark~4]{Zattoni-2013}, the selection of the control inputs has implied an increase 
in the relative degree from $2$ to $3$.
%
%
%
%
%
\section{Conclusion}
Three numerical examples have been presented with the aim of illustrating the
geometric techniques for accomplishing zero cancellation in linear multivariable
systems devised in \cite{Zattoni-2013}. Benefits of the proposed methodology
compared with those available in the literature have been shown.
\bibliographystyle{IEEEtran}
\bibliography{asjc2013arXiv}
\end{document}